\documentstyle[12pt,epsf]{article}
\begin{document}
\baselineskip 0.7cm

\newcommand{\gsim}{ \mathop{}_{\textstyle \sim}^{\textstyle >} }
\newcommand{\lsim}{ \mathop{}_{\textstyle \sim}^{\textstyle <} }
\newcommand{\vev}[1]{ \left\langle {#1} \right\rangle }
\newcommand{\yrs}{ {\rm years} }
\newcommand{\EV}{ {\rm eV} }
\newcommand{\KEV}{ {\rm keV} }
\newcommand{\MEV}{ {\rm MeV} }
\newcommand{\GEV}{ {\rm GeV} }
\newcommand{\TEV}{ {\rm TeV} }
\newcommand{\mgut}{M_{GUT}}
\newcommand{\mint}{M_{I}}
\newcommand{\mgra}{M_{3/2}}
\newcommand{\mgy}{m_{G1}}
\newcommand{\mgl}{m_{G2}}
\newcommand{\mgc}{m_{G3}}
\newcommand{\nuR}{\nu_{R}}
\newcommand{\e}{{\rm e}}
\renewcommand{\thefootnote}{\fnsymbol{footnote}}
\setcounter{footnote}{1}

\begin{titlepage}

\begin{flushright}
UT-837
\end{flushright}

\vskip 0.35cm
\begin{center}
{\large \bf  Long-Lived Superheavy Particles \\
             in Dynamical Supersymmetry-Breaking Models \\
             in Supergravity}
\vskip 1.2cm
K.~Hamaguchi$^{1}$, Izawa~K.-I.$^{1}$, Yasunori Nomura$^{1}$ 
and T.~Yanagida$^{1,2}$

\vskip 0.4cm

$^{1}$ {\it Department of Physics, University of Tokyo, 
            Tokyo 113-0033, Japan}
\\
$^{2}$ {\it Research Center for the Early Universe, University of Tokyo,\\
         Tokyo 113-0033, Japan}

\vskip 1.5cm

\abstract{Superheavy particles of masses $\simeq 10^{13}-10^{14}~\GEV$
 with lifetimes $\simeq 10^{10}-10^{22}~\yrs$ are very interesting,
 since their decays may account for the ultra-high energy (UHE) cosmic
 rays discovered beyond the Greisen-Zatsepin-Kuzmin cut-off energy 
 $E \sim 5 \times 10^{10}~\GEV$.
 We show that the presence of such long-lived superheavy particles is a
 generic prediction of QCD-like SU($N_c$) gauge theories with $N_f$
 flavors of quarks and antiquarks and the large number of colors $N_c$.
 We construct explicit models based on supersymmetric SU($N_c$) gauge
 theories and show that if the dynamical scale 
 $\Lambda \simeq 10^{13}-10^{14}~\GEV$ and $N_c = 6-10$ the lightest
 composite baryons have the desired masses and lifetimes to explain the
 UHE cosmic rays.
 Interesting is that in these models the gaugino condensation
 necessarily occurs and hence these models may play a role of so-called
 hidden sector for supersymmetry breaking in supergravity.}

\end{center}
\end{titlepage}

\renewcommand{\thefootnote}{\arabic{footnote}}
\setcounter{footnote}{0}

%
%
%
%

\section{Introduction}

In QCD-like SU($N_c$) gauge theories with $N_f$ flavors of quarks $Q$
and antiquarks $\bar{Q}$, any charges (like baryon number) associated
with conserved vector currents are not spontaneously broken \cite{VW}.
Thus, the lightest bound states of $Q$'s or $\bar{Q}$'s carrying
nonvanishing baryon numbers are almost stable and they will decay into
the ordinary quarks and leptons through some baryon-number violating
nonrenormalizable operators suppressed by the gravitational scale 
$M_* \simeq 2.4 \times 10^{18}~\GEV$.
If the dynamical scale $\Lambda$ of the SU($N_c$) gauge interactions 
is well below the gravitational scale and $N_c$ is sufficiently large,
the lifetimes of the superheavy baryons may be longer than the age of
the present universe.
Therefore, the presence of long-lived superheavy baryons is a generic
prediction of QCD-like gauge theories for a certain parameter region of
$N_c$ and $\Lambda$.

However, any (quasi-)stable particles much heavier than $O(1)~\TEV$ are
cosmologically dangerous, since they would easily overclose the universe 
if they were once in the thermal equilibrium \cite{GK-PRL}.
One may usually invoke some inflationary stage in the universe's
evolution to dilute the number density of such superheavy $X$ particles.
If the reheating temperature after the inflation is much lower than the
masses of $X$ particles one may completely neglect the thermal
production of $X$ particles.
It has been, however, suggested \cite{CKR_production, KT_production}
that if the $X$-particle masses $m_X$ are of order of the Hubble
constant $H$ at the final epoch of inflation, gravitational interactions
may give nonnegligible contributions to the $X$-particle production just
after the end of the inflation.
The numerical calculation in Ref.~\cite{CKR_production}, in fact, shows
that when $m_X \simeq (0.04-2)\times H$ a suitable amount of the $X$
particles is produced to form a part of the dark matter in the present
universe.
It is remarkable that the decays of such $X$ particles will generate
significant effects on the spectrum of high energy cosmic ray if the
lifetimes of the $X$ particles are of order of the age of the present
universe.

Several events of the ultra-high energy (UHE) cosmic rays 
\cite{UHE_CR1, UHE_CR2, UHE_CR3} have been recently observed beyond
the Greisen-Zatsepin-Kuzmin (GZK) bound 
$E \sim 5 \times 10^{10}~\GEV$ \cite{GZK_cutoff}.
These are naturally explained \cite{UHE_SH1, UHE_SH2} by the decay
products of superheavy $X$ particle of mass 
$m_X \simeq 10^{13}-10^{14}~\GEV$\footnote{
There is an analysis which suggests $m_X \simeq 10^{12}~\GEV$
\cite{BS}.}
with lifetime 
$\tau_X \simeq 10^{10}-10^{22}~\yrs$\footnote{
The required lifetime may be accounted for by imposing discrete gauge
symmetries even if the $X$ are elementary particles
\cite{discrete_gauge}.}
if its energy density $\rho_X$ lies in the range
\begin{eqnarray}
  \rho_X / \rho_c \simeq 10^{-12}-1.
\end{eqnarray}
Here, $\rho_c \simeq 8.1 h^2 \times 10^{-47}~\GEV^4$ with 
$h \simeq 0.5-1.0$ is the critical density of the present universe.
Since the required window of the energy density $\rho_X$ is very wide,
this scenario seems very plausible and attractive.

In this paper, we construct explicit models based on supersymmetric
(SUSY) SU($N_c$) gauge theories with $N_f$ flavors of quarks $Q$ and
antiquarks $\bar{Q}$, in which the lightest baryons $B$ and antibaryons
$\bar{B}$ have the desired mass and lifetime, that is, 
$m_B = m_{\bar{B}} \simeq 10^{13}-10^{14}~\GEV$ and 
$\tau_B = \tau_{\bar{B}} \simeq 10^{10}-10^{22}~\yrs$.
In these models the gaugino condensation necessarily occurs and hence
the models may play a role of so-called hidden sector for SUSY breaking
in supergravity \cite{gaugino_cond}.
Thus, the long-lived superheavy $B$ and $\bar{B}$ are regarded as
byproducts\footnote{
A similar idea has been considered in connection with string theory
\cite{UHE_M}.}
of the hidden sector gauge theories for dynamical SUSY
breaking.\footnote{
We may consider non-SUSY SU($N_c$) gauge theories which cause dynamical
breaking of the Peccei-Quinn symmetry at $\Lambda \simeq 10^{13}~\GEV$
\cite{axion}.
In these models we may have naturally quasi-stable baryons of masses
$\sim 10^{13}-10^{14}~\GEV$, which have the required lifetimes 
$\tau_B \simeq 10^{10}-10^{22}~\yrs$ for $N_c \simeq 5, 6$.
The main decay mode of such baryons will be 
$B \rightarrow l+$Higgs boson or $2\times$Higgs bosons.}

It should be noted here that in contrast to the non-SUSY case, the SUSY
QCD-like gauge theories may yield baryon-number violating vacua due to
the presence of scalar quarks.
In these vacua we have no longer quasi-stable baryons.
However, if quarks $Q$ and antiquarks $\bar{Q}$ have SUSY-invariant
masses, the unwanted baryon-number violating vacua disappear as shown in 
Ref.~\cite{moduli_space}.
Thus, we consider the SUSY QCD-like gauge theries with massive quarks
throughout this paper.

In section 2, we briefly discuss vacua of SUSY SU($N_c$) gauge theories
with $N_f$ pairs of massive quarks and antiquarks.
We restrict our discussion to the case of $N_f=N_c+1$ and show that
there is a unique SUSY-invariant vacuum preserving the baryon-number
conservation.
Thus, we always have stable baryons and antibaryons in this theory.
In section 3, we extend the above model to the supergravity, in which we 
introduce nonrenormalizable operators.
We show that possible baryon-number violating operators induce
spontaneous breakdown of the baryon-number conservation and the
baryon-meson mixings occur.
Owing to the baryon-number violating effects even the lightest baryons
are no longer stable.
However, we see that the lifetimes of the baryons can be chosen as in
the required range to account for the UHE cosmic rays by taking 
$N_c = 8,9$ and $10$.
We also show that the gaugino condensation necessarily occurs and the
SUSY may be broken in the dilaton stabilized vacua.
In section 4, we argue that the lifetimes of the baryons become longer 
if the baryons carry nonvanishing charges of some extra symmetries.
As for such symmetries we adopt the matter parity ${\bf Z}_2$ or the
discrete baryon parity ${\bf Z}_3$, since they are often used to
guarantee the stability of usual proton \cite{proton}.
In these cases we find the desired lifetimes are obtained for somewhat
smaller $N_c = 6-9$.
The last section 5 is devoted to discussion and conclusions.

\section{Supersymmetric QCD with Massive Quarks}

Let us consider SUSY SU($N_c$) gauge theories with $N_f$ flavors of
quarks $Q_a^i$ and antiquarks $\bar{Q}^a_{\bar{\imath}}$, where 
$a = 1,\cdots,N_c$ and $i,\bar{\imath}=1,\cdots,N_f$.
We omit the color index $a$, hereafter.
We neglect the mass term for $Q^i$ and $\bar{Q}_{\bar{\imath}}$ for the
time being.
Then, we have a global 
SU($N_f$)$\times$SU($N_f$)$\times$U(1)$_V \times$U(1)$_R$ symmetry.
We restrict our consideration to the case $N_f = N_c+1$, since the
dynamics is the clearest in this case.

For $N_f = N_c+1$, the low energy physics is described by 
canonically-normalized gauge invariant composite fields, mesons 
$M^i_{\bar{\jmath}} \simeq Q^i\bar{Q}_{\bar{\jmath}}/\Lambda$, baryons 
$B_i \simeq \epsilon_{ijk\cdots l}Q^jQ^k\cdots Q^l/\Lambda^{N_c-1}$ and
antibaryons $\bar{B}^{\bar{\imath}} \simeq 
\epsilon^{\bar{\imath}\bar{\jmath}\bar{k}\cdots\bar{l}}
\bar{Q}_{\bar{\jmath}}\bar{Q}_{\bar{k}}\cdots\bar{Q}_{\bar{l}}
/\Lambda^{N_c-1}$ \cite{SUSY_QCD}.
The dynamically generated superpotential is given by
\begin{eqnarray}
  W_{\rm dyn} = B_i M^i_{\bar{\jmath}} \bar{B}^{\bar{\jmath}} 
      - \frac{1}{\Lambda^{N_c-2}} {\rm det}M.
\end{eqnarray}
The mesons $M^i_{\bar{\jmath}}$, baryons $B_i$ and antibaryons
$\bar{B}^{\bar{\imath}}$ are all massless to satisfy the 't Hooft
anomaly matching conditions \cite{SUSY_QCD}.

Now we introduce the mass term for quarks $Q^i$ and antiquarks
$\bar{Q}_{\bar{\imath}}$.
Then, the total effective superpotential is given by
\begin{eqnarray}
  W = B_i M^i_{\bar{\jmath}} \bar{B}^{\bar{\jmath}} 
      - \frac{1}{\Lambda^{N_c-2}} {\rm det}M
      + m_i^{\bar{\jmath}} \Lambda M^i_{\bar{\jmath}}.
\end{eqnarray}
It is a straightforward task to see a SUSY invariant vacuum:
\begin{eqnarray}
  && \vev{M^i_{\bar{\jmath}}} = \Lambda^{\frac{N_c-1}{N_c}}
     ({\rm det}m)^{\frac{1}{N_c}} (m^{-1})^i_{\bar{\jmath}}, \\
  && \vev{B_i} = \vev{\bar{B}^{\bar{\imath}}} = 0.
\end{eqnarray}
In this vacuum the mesons $M^i_{\bar{\jmath}}$, baryons $B_i$ and
antibaryons $\bar{B}^{\bar{\imath}}$ have the following mass terms
\begin{eqnarray}
  W_{\rm mass} &=& \Lambda^{\frac{N_c-1}{N_c}}
       ({\rm det}m)^{\frac{1}{N_c}} (m^{-1})^i_{\bar{\jmath}}
       B_i \bar{B}^{\bar{\jmath}} \nonumber\\
  && - \frac{1}{2} \Lambda^{\frac{1}{N_c}}
       ({\rm det}m)^{\frac{-1}{N_c}} 
       (m^{\bar{\jmath}}_i m^{\bar{l}}_k - m^{\bar{\jmath}}_k m^{\bar{l}}_i)
       M^i_{\bar{\jmath}} M^k_{\bar{l}}.
\end{eqnarray}
Thus, we obtain the masses for these composite fields as
\begin{eqnarray}
  m_B &\simeq& (m\Lambda^{N_c-1})^{\frac{1}{N_c}}, \\
  m_M &\simeq& (m^{N_c-1}\Lambda)^{\frac{1}{N_c}},
\label{diagonal_mass}
\end{eqnarray}
where we have assumed a common mass 
$m_i^{\bar{\jmath}} = m \delta_i^{\bar{\jmath}}$, for simplicity.
We will identify these composite baryons $B_i$ and antibaryons
$\bar{B}^{\bar{\imath}}$ with the long-lived superheavy $X$ particle
introduced to explain the UHE cosmic rays discovered beyond the GZK
bound.
Therefore, we take
\begin{eqnarray}
  m_B \simeq (m\Lambda^{N_c-1})^{\frac{1}{N_c}}
      \simeq 10^{13}-10^{14}~\GEV.
\label{X_mass}
\end{eqnarray}

Using the Konishi anomaly relation \cite{K-A} we determine the gaugino
condensation as 
\begin{eqnarray}
\begin{array}{ccccc}
  \vev{\lambda\lambda} &=& \frac{1}{N_c+1}m^{\bar{\jmath}}_i
      \vev{Q^i\bar{Q}_{\bar{\jmath}}} 
      &=& \Lambda^{\frac{2N_c-1}{N_c}} ({\rm det}m)^{\frac{1}{N_c}} \\ \\
  &\simeq& (m^{N_c+1}\Lambda^{2N_c-1})^{\frac{1}{N_c}}. &&
\end{array}
\label{gaugino}
\end{eqnarray}
This condensation may give a dominant contribution to the SUSY breaking 
in supergravity.

\section{Extension to the Supergravity}

We now extend the above model to the supergravity.
So far we have considered only renormalizable interactions, but in the
framework of supergravity it is quite natural to introduce
nonrenormalizable interactions.
Namely, we introduce in general\footnote{
There exist also baryon-number violating nonrenormalizable operators in
the K\"{a}hler potential.
However, they are negligible compared with the baryon-number violating
operators in the superpotential.}
\begin{eqnarray}
  W_{\rm tree} = m_i^{\bar{\jmath}} Q^i \bar{Q}_{\bar{\jmath}}
      + \frac{b^i}{M_*^{N_c-3}}\epsilon_{ijk\cdots l}Q^jQ^k\cdots Q^l
      + \frac{\bar{b}_{\bar{\imath}}}{M_*^{N_c-3}}
        \epsilon^{\bar{\imath}\bar{\jmath}\bar{k}\cdots\bar{l}}
        \bar{Q}_{\bar{\jmath}}\bar{Q}_{\bar{k}}\cdots\bar{Q}_{\bar{l}},
\end{eqnarray}
with $b_i, \bar{b}_{\bar{\imath}} = O(1)$.
We have omitted the other nonrenormalizable terms which are irrelevant
for our purposes here.
Then, the total effective superpotential is given by
\begin{eqnarray}
  W &=& W_{\rm dyn} + W_{\rm tree} \nonumber\\
    &=& B_i M^i_{\bar{\jmath}} \bar{B}^{\bar{\jmath}} 
       - \frac{1}{\Lambda^{N_c-2}} {\rm det}M \nonumber\\
    && + m_i^{\bar{\jmath}} \Lambda M^i_{\bar{\jmath}}
       + b^i \left( \frac{\Lambda}{M_*} \right)^{N_c-3} \Lambda^2 B_i
       + \bar{b}_{\bar{\imath}} \left( \frac{\Lambda}{M_*} \right)^{N_c-3} 
         \Lambda^2 \bar{B}^{\bar{\imath}}.
\label{totalpotential}
\end{eqnarray}
It is a straightforward task to see a SUSY invariant vacuum:
\begin{eqnarray}
  \vev{M^i_{\bar{\jmath}}} &=& \Lambda^{\frac{N_c-1}{N_c}}
     ({\rm det}m)^{\frac{1}{N_c}} (m^{-1})^i_{\bar{\jmath}}, \\
  \vev{B_i} &=& -\bar{b}_{\bar{\jmath}}
     \left( \frac{\Lambda}{M_*} \right)^{N_c-3}
     \Lambda^{\frac{N_c+1}{N_c}}({\rm det}m)^{\frac{-1}{N_c}}
     m_i^{\bar{\jmath}}, 
\label{baryon-cond-1}\\
  \vev{\bar{B}^{\bar{\imath}}} &=& -b^j
     \left( \frac{\Lambda}{M_*} \right)^{N_c-3}
     \Lambda^{\frac{N_c+1}{N_c}}({\rm det}m)^{\frac{-1}{N_c}}
     m_j^{\bar{\imath}},
\label{baryon-cond-2}
\end{eqnarray}
up to the leading order in $\Lambda/M_*$.
In this vacuum the mesons $M^i_{\bar{\jmath}}$, baryons $B_i$ and
antibaryons $\bar{B}^{\bar{\imath}}$ have the following mass terms
\begin{eqnarray}
  W_{\rm mass} &=& \Lambda^{\frac{N_c-1}{N_c}}
       ({\rm det}m)^{\frac{1}{N_c}} (m^{-1})^i_{\bar{\jmath}}
       B_i \bar{B}^{\bar{\jmath}} \nonumber\\
  && - \frac{1}{2} \Lambda^{\frac{1}{N_c}}
       ({\rm det}m)^{\frac{-1}{N_c}} 
       (m^{\bar{\jmath}}_i m^{\bar{l}}_k - m^{\bar{\jmath}}_k m^{\bar{l}}_i)
       M^i_{\bar{\jmath}} M^k_{\bar{l}} \nonumber\\
  && - b^k \left( \frac{\Lambda}{M_*} \right)^{N_c-3}
     \Lambda^{\frac{N_c+1}{N_c}}({\rm det}m)^{\frac{-1}{N_c}}
     m_k^{\bar{\jmath}} B_i M^i_{\bar{\jmath}} \nonumber\\
  && - \bar{b}_{\bar{k}} \left( \frac{\Lambda}{M_*} \right)^{N_c-3}
     \Lambda^{\frac{N_c+1}{N_c}}({\rm det}m)^{\frac{-1}{N_c}}
     m_i^{\bar{k}} M^i_{\bar{\jmath}} \bar{B}^{\bar{\jmath}}.
\label{composite_mass}
\end{eqnarray}
Thus, we obtain mixing masses between the composite meson and
baryon fields as
\begin{eqnarray}
  m_{BM} &\simeq& \left( \frac{\Lambda}{M_*} \right)^{N_c-3}
                  (m^{-1}\Lambda^{N_c+1})^{\frac{1}{N_c}}.
\end{eqnarray}
The diagonal masses $m_B$ for baryons and $m_M$ for mesons are the same
as in Eq.~(\ref{diagonal_mass}) up to the leading order in
$\Lambda/M_*$.
Notice that as long as $\Lambda \ll M_*$ and $N_c \gg 3$ the mixings
between mesons and baryons are very small.\footnote{
The baryon-number condensations in 
Eqs.~(\ref{baryon-cond-1}, \ref{baryon-cond-2}) induce kinetic mixings
between the meson and baryon fields in the K\"{a}hler potential, which
give the same-order effects as those discussed in the text.
We neglect them, for simplicity, since they do not affect our main
conclusions.}

Let us now discuss the decay of these $B_i$ and
$\bar{B}^{\bar{\imath}}$.
When the masses $m_M$ for the mesons are all larger than the half of
those of $B_i$ and $\bar{B}^{\bar{\imath}}$, these composite baryons
should decay directly into ordinary light particles including quarks and 
leptons through the mixing terms in Eq.~(\ref{composite_mass}) together
with the following nonrenormalizable operator:
\begin{eqnarray}
  W = \frac{f}{M_*} Q\bar{Q} H\bar{H}.
\label{meson_decay}
\end{eqnarray}
The lifetimes of $B_i$ and $\bar{B}^{\bar{\imath}}$ are determined as
\begin{eqnarray}
  \tau_B \simeq \frac{1}{f^2} 
         \left( \frac{M_*}{\Lambda} \right)^{2(N_c-2)}
         (m^{-3}\Lambda^{N_c+3})^{\frac{-1}{N_c}},
\label{lifetime-B_1}
\end{eqnarray}
which should be taken $\simeq 10^{10}-10^{22}~\yrs$ to account for the
UHE cosmic rays.

When $2m_{M} < m_{B}$, we have new decay channels 
$B_i\,(\bar{B}^{\bar{\imath}}) \rightarrow 2M^i_{\bar{\jmath}}$.
Since the mesons $M^i_{\bar{\jmath}}$ decay into ordinary light
particles very quickly through the interactions Eq.~(\ref{meson_decay}),
the lifetimes of $B_i$ and $\bar{B}^{\bar{\imath}}$ decaying into the
ordinary light particles are determined by the $2M^i_{\bar{\jmath}}$ 
decay channels which are given by
\begin{eqnarray}
  \tau_B \simeq \left( \frac{M_*}{\Lambda} \right)^{2(N_c-3)}
         (m^{2N_c-7}\Lambda^{-N_c+7})^{\frac{-1}{N_c}}.
\end{eqnarray}
These are somewhat shorter than the previous lifetimes
Eq.~(\ref{lifetime-B_1}).
Thus, we conclude
\begin{eqnarray}
  \left\{ \matrix{
  \tau_B^{-1} &\simeq& f^2 
      \left( \frac{\Lambda}{M_*} \right)^{2(N_c-2)}
      (m^{-3}\Lambda^{N_c+3})^{\frac{1}{N_c}} \cr
      &\simeq& 10^{-54}-10^{-42}~\GEV 
      &({\rm for}\,\, 2m_{M} > m_{B}), \cr \cr
  \tau_B^{-1} &\simeq& \left( \frac{\Lambda}{M_*} \right)^{2(N_c-3)}
      (m^{2N_c-7}\Lambda^{-N_c+7})^{\frac{1}{N_c}} \cr
      &\simeq& 10^{-54}-10^{-42}~\GEV
      &({\rm for}\,\, 2m_{M} < m_{B}), \cr}
  \right.
\label{X_lifetime}
\end{eqnarray}
to have the required lifetime $\tau_B \simeq 10^{10}-10^{22}~\yrs$.

Now we are at the point to discuss the gaugino condensation
Eq.~(\ref{gaugino}).
It is well known that this gaugino condensation may induce SUSY breaking 
together with the dilaton field stabilization in supergravity
\cite{gaugino_cond}.\footnote{
The SUSY-breaking effects do not change the order of magnitude of
$\vev{\lambda\lambda}$ in Eq.~(\ref{gaugino}).}
Assuming the gravitino mass $m_{3/2} \simeq 1~\TEV$ we determine the
gaugino condensation scale 
$\vev{\lambda\lambda}^{1/3} \simeq 10^{13}~\GEV$.
From the constraints Eqs.~(\ref{X_mass}, \ref{gaugino},
\ref{X_lifetime}) we obtain the following relations:
\begin{eqnarray}
  \left\{ \matrix{
  \frac{1}{f^2} M_*^{2(N_c-2)} 
      \vev{\lambda\lambda}^{\frac{2N_c}{N_c-2}} \tau_B^{-1}
      &=& m_B^{\frac{2N_c^2-N_c+6}{N_c-2}}
      &({\rm for}\,\, 2m_{M} > m_{B}), \cr
  M_*^{2(N_c-3)} \vev{\lambda\lambda}^{\frac{2}{N_c-2}} \tau_B^{-1}
      &=& m_B^{\frac{2N_c^2-9N_c+16}{N_c-2}}
      &({\rm for}\,\, 2m_{M} < m_{B}). \cr}
  \right.
\end{eqnarray}
These relations are consistent with the required values of $m_B$,
$\tau_B$ and $\vev{\lambda\lambda}$ only when the numbers of colors
$N_c$ are
\begin{eqnarray}
  \left\{ \matrix{
  N_c &=& 8, 9, 10  &({\rm for}\,\, 2m_{M} > m_{B}), \cr
  N_c &=& 9, 10     &({\rm for}\,\, 2m_{M} < m_{B}). \cr}
  \right.
\end{eqnarray}
Here, we have assumed the coupling constant $f \simeq 1$.
Then, $\Lambda$ and $m$ are determined as
\begin{eqnarray}
  \Lambda &\simeq& 10^{13.0}-10^{14.5}~\GEV, 
   \label{Lambda-res}\\
  m       &\simeq& 10^{10.5}-10^{13.0}~\GEV.
   \label{m-res}
\end{eqnarray}
Note that these numerical values yield too large a mass term for the
Higgs doublets in Eq.~(\ref{meson_decay}).
We postpone the discussion on this point to the final section.

\section{Models with Discrete Gauge Symmetries}

In the previous section we find that the desired lifetimes of the
baryons are obtained if $N_c = 8, 9$ and $10$ without invoking 
any extra symmetries.
In this section, we impose the matter parity ${\bf Z}_2$ or the baryon
parity ${\bf Z}_3$ on our model, since these discrete gauge symmetries
are often used to suppress very rapid decays of the usual
proton.\footnote{
In the case that neutrinos acquire Majorana masses through operators
$W = (1/M_R)llHH$, the anomaly-free discrete gauge symmetries are only
the matter parity ${\bf Z}_2$ and the baryon parity ${\bf Z}_3$
\cite{discrete_gauge}.}
The charges for the minimal SUSY standard-model (MSSM) particles under
the matter parity ${\bf Z}_2$ and the baryon parity ${\bf Z}_3$ are
given in Table \ref{table_charges}.
\begin{table}
 \begin{center}
  \begin{tabular}{|c|c c c c c c c|}\hline
               & $q$ & $\bar{u}$ & $\bar{d}$ & $l$ & $\bar{e}$
   & $H$ & $\bar{H}$ \\ \hline
   ${\bf Z}_2$ & $0$ & $1$       & $1$       & $0$ & $1$
   & $1$ & $1$   \\
   ${\bf Z}_3$ & $0$ & $2$       & $1$       & $2$ & $2$
   & $1$ & $2$   \\ \hline
  \end{tabular}
 \end{center}
 \caption{Charges for the MSSM particles under the matter parity
 ${\bf Z}_2$ and the baryon parity ${\bf Z}_3$.
 $q,\bar{u},\bar{d},l$ and $\bar{e}$ denote $SU(2)_L$-doublet quark,
 up-type antiquark, down-type antiquark, $SU(2)_L$-doublet lepton and 
 charged antilepton chiral multiplets.
 $H$ and $\bar{H}$ are chiral multiplets for Higgs doublets.
 }
 \label{table_charges}
\end{table}
If the composite baryons do not carry nonvanishing charges of the 
${\bf Z}_2$ and ${\bf Z}_3$, the analyses are the same as in the
previous section.
If these composite baryons have nontrivial ${\bf Z}_2$ or ${\bf Z}_3$
charges, however, there are no mixing mass terms between mesons and
baryons, since the linear terms of $B_i$ and
$\bar{B}^{\bar{\imath}}$ in Eq.~(\ref{totalpotential}) are forbidden
({\it i.e.} $b^i = \bar{b}_{\bar{\imath}} = 0$)
and hence $\vev{B_i} = \vev{\bar{B}^{\bar{\imath}}} = 0$.
Thus, the decay channels of $B_i$ and $\bar{B}^{\bar{\imath}}$ are
different from those in the previous section.
In this section, we discuss the lifetimes of the baryons with the
discrete ${\bf Z}_2$ or ${\bf Z}_3$ \cite{discrete_gauge}.

First, we consider the case of the matter parity ${\bf Z}_2$.
The charges for the $B_i$ and $\bar{B}^{\bar{\imath}}$ must be opposite
for quarks $Q^i$ and antiquarks $\bar{Q}_{\bar{\imath}}$ to have
invariant masses.
Thus, we suppose that both $B_i$ and $\bar{B}^{\bar{\imath}}$ are odd under
the ${\bf Z}_2$.
Then, the lowest dimensional operators which cause decays of the
composite baryons are
\begin{eqnarray}
  W = \frac{1}{M_*^{N_c-1}}QQ\cdots Q\,lH
   + \frac{1}{M_*^{N_c-1}}\bar{Q}\bar{Q}\cdots\bar{Q}\,lH,
\end{eqnarray}
where $Q$'s and $\bar{Q}$'s, for example, carry ${\bf Z}_2$ charges 
$(2r+1)/N_c \,\,(r\in {\bf Z})$ and $-(2r+1)/N_c$,
respectively.\footnote{
The ${\bf Z}_2$ is anomaly free.}
Then, the lifetimes of $B_i$ and 
$\bar{B}^{\bar{\imath}}$ are determined as
\begin{eqnarray}
 \tau_B\simeq \left( \frac{M_*}{\Lambda} \right)^{2(N_c-1)}
  (m\Lambda^{N_c-1})^{\frac{-1}{N_c}}.
\label{X-lifetime-z2}
\end{eqnarray}

Next, let us turn to the case of the baryon parity ${\bf Z}_3$.
We suppose that the $B_i$ carry $+1$ and the
$\bar{B}^{\bar{\imath}}$ carry $-1$ of the ${\bf Z}_3$ charges.
Then, the lightest baryons decay into the MSSM particles through the
following operator:
\begin{eqnarray}
 W = \frac{1}{M_*^{N_c}}\bar{Q}\bar{Q}\cdots\bar{Q}\,\bar{u}\bar{d}\bar{d},
\end{eqnarray}
where $Q$'s and $\bar{Q}$'s carry ${\bf Z}_3$ charges $(3r+1)/N_c$ and
$-(3r+1)/N_c$, respectively.\footnote{
The ${\bf Z}_3$ is anomaly free.}
The lifetimes of $B_i$ and $\bar{B}^{\bar{\imath}}$ are given by
\begin{eqnarray}
 \tau_B\simeq \left( \frac{M_*}{\Lambda} \right)^{2N_c}
  (m^3\Lambda^{N_c-3})^{\frac{-1}{N_c}}.
\label{X-lifetime-z3}
\end{eqnarray}
Note that even if we assign the ${\bf Z}_3$ charge $-1$ for $B_i$ and
$+1$ for $\bar{B}^{\bar{\imath}}$, the lightest baryons decay through
the operator
\begin{eqnarray}
 W = \frac{1}{M_*^{N_c}}QQ\cdots Q\,\bar{u}\bar{d}\bar{d},
\end{eqnarray}
so that the lifetimes are the same as in Eq.~(\ref{X-lifetime-z3}).

From the constraints Eqs.~(\ref{X_mass}, \ref{gaugino},
\ref{X-lifetime-z2}, \ref{X-lifetime-z3}) we obtain the following
relations:
\begin{eqnarray}
 \left\{ \matrix{
  M_*^{2(N_c-1)} \vev{\lambda\lambda}^{\frac{2(N_c-1)}{N_c-2}} \tau_B^{-1}
  &=& m_B^{\frac{2N_c^2+N_c-4}{N_c-2}}
  &({\rm for}\,\,{\bf Z}_2), \cr
  M_*^{2N_c} \vev{\lambda\lambda}^{\frac{2(N_c-1)}{N_c-2}} \tau_B^{-1}
  &=& m_B^{\frac{2N_c^2+3N_c-8}{N_c-2}}
  &({\rm for}\,\,{\bf Z}_3), \cr}
  \right.
\end{eqnarray}
which are consisitent with the required values of $m_B$, $\tau_B$ and
$\vev{\lambda\lambda}$ only when the numbers of colors $N_c$ are
\begin{eqnarray}
  \left\{ \matrix{
  N_c &=& 7, 8, 9, &({\rm for}\,\,{\bf Z}_2), \cr
  N_c &=& 6, 7, 8  &({\rm for}\,\,{\bf Z}_3). \cr}
  \right.
\end{eqnarray}
Therefore, when the $B_i$ and $\bar{B}^{\bar{\imath}}$ carry
nonvanishing charges of the matter parity ${\bf Z}_2$ or the baryon
parity ${\bf Z}_3$, the required numbers of colors $N_c$ to obtain the
desired lifetimes are smaller than those in the case without 
${\bf Z}_2$ or ${\bf Z}_3$.
Notice that $\Lambda$ and $m$ are almost the same as in
Eqs.~(\ref{Lambda-res}, \ref{m-res}) inspite of the change of $N_c$.

\section{Discussion and Conclusions}

In this paper we have constructed explicit models based on SUSY
SU($N_c$) gauge theories with $N_f$ flavors of quarks and antiquarks
which naturally accommodate the superheavy composite baryons $B$ and
$\bar{B}$ introduced to account for the UHE cosmic rays beyond the GZK
bound.
The models contain three crucial parameters, the quark mass $m$, the
dynamical scale $\Lambda$ and the number of colors $N_c$.
The number of flavors $N_f$ is fixed as $N_f = N_c + 1$, for simplicity.
In these models the gaugino condensation $\vev{\lambda\lambda}$ always
occurs, which may cause the SUSY breaking in supergravity.
Assuming $\vev{\lambda\lambda}^{1/3}\simeq 10^{13}~\GEV$
(for $m_{3/2}\simeq 1~\TEV$)
and the desired properties for the composite baryons
($m_B = m_{\bar{B}} \simeq 10^{13}-10^{14}~\GEV$ and 
$\tau_B = \tau_{\bar{B}} \simeq 10^{10}-10^{22}~\yrs$),
we have obtained $m\simeq 10^{10.5}-10^{13.0}\GEV$,
$\Lambda\simeq 10^{13.0}-10^{14.5}\GEV$ and $N_c = 6 - 10$.
Namely, we have found that the required long lifetimes of superheavy $B$ 
and $\bar{B}$ are naturally explained in SUSY QCD-like gauge theories
with large number $N_c$ of color degrees of freedom.
Although we have restricted our analysis only to the case of 
$N_f = N_c + 1$, it is possible to consider other cases.

Several comments are in order.
We should mention first the so-called $\mu$ term problem.
Owing to the $Q\bar{Q}$ condensation, the Higgs doublets seem to have an 
invariant mass term $\mu H\bar{H}$, where
$\mu\simeq f\vev{Q\bar{Q}}/M_*\simeq (10^{8}-10^{10})\times f~\GEV$.
Thus, in order to induce a correct vacuum-expectation values for the
Higgs doublets, a negative invariant mass of order $10^{8}-10^{10}~\GEV$
must be introduced to cancell the unwanted large mass
$f\vev{Q\bar{Q}}/M_*$.
Alternatively, one can solve this problem by putting the coupling
constant $f$ of the operator $(1/M_*)Q\bar{Q}H\bar{H}$ very small 
$f \simeq 10^{-5}-10^{-7}$.
We see that even if it is the case, the obtained $\Lambda$ and $m$ are
almost the same as in Eqs.~(\ref{Lambda-res}, \ref{m-res}), and our
conclusion does not change much.\footnote{
For the case of $2m_M > m_B$, if we take $f \simeq 10^{-5}-10^{-7}$ the
dominant operators contributing to the baryon decays are not those in
Eq.~(\ref{meson_decay}) but direct decay operators given by 
\begin{eqnarray}
 W = \frac{1}{M_*^{N_c-1}}QQ\cdots Q\,H\bar{H}
   + \frac{1}{M_*^{N_c-1}}\bar{Q}\bar{Q}\cdots\bar{Q}\,H\bar{H},
\label{direct_decay}
\end{eqnarray}
which give the desired number of colors $N_c = 7, 8, 9$ instead of 
$N_c = 8, 9, 10$ obtained in the text.
On the other hand, the lifetimes are independent of the value of $f$ for
the case of $2m_M < m_B$.}
The small value of $f$ will be understood by some axial symmetries,
which may also explain the small mass $m$ for $Q^i$ and
$\bar{Q}_{\bar{\imath}}$.\footnote{
$R$ symmetry may be an example in which $H\bar{H}$ has $R$-charge zero.
We naturally obtain the terms $mQ\bar{Q} (1+f'H\bar{H}/M_*^2)$ in
the superpotential with a coupling $f'$ of order one, which may yield an 
appropriate $\mu$ term.
The small mass $m$ is regarded as a breaking term of the $R$ symmetry.
To have unsuppressed baryon-number violating operators in
Eq.~(\ref{direct_decay}), for example, we assume that both of $Q^i$ and
$\bar{Q}_{\bar{\imath}}$ carry $R$-charge $2/N_c$.}

Our model is also applicable to the gauge-mediated SUSY breaking
scenario \cite{gauge-med}.
If the gaugino condensation causes SUSY breaking partially, the induced
gravitino mass ``$m_{3/2}$'' must be smaller than $1~\GEV$ in order to
suppress dangerous flavor-changing neutral currents sufficiently.
We find that the desired composite baryons are obtained if 
$m \simeq 10^{2.0}-10^{9.7}~\GEV$, 
$\Lambda \simeq 10^{13.3}-10^{16.0}~\GEV$ and $N_c = 6-11$ for 
``$m_{3/2}$''$\simeq 100~\KEV - 1~\GEV$.

Finally, we should comment on the gauge coupling constant $\alpha_c$ of
the SUSY SU($N_c$) gauge theories considered in this paper.
With $N_c = 6 - 10$ and $N_f = N_c+1$ the solution to the one-loop
renormalization group equation for the gauge coupling constant
$\alpha_c$ is given by 
\begin{eqnarray}
  \alpha_c^{-1}(\Lambda) = \alpha_c^{-1}(M_*) 
    + \frac{2N_c-1}{2\pi} \ln \left( \frac{\Lambda}{M_*} \right).
\end{eqnarray}
Using $\alpha_c(\Lambda) \simeq \infty$ we get 
$\alpha_c(M_*) \simeq 0.03 - 0.05$ at the gravitational scale $M_*$.
It is interesting that the value of $\alpha_c(M_*)$ is roughly
consistent with the hypothesis of unification with the standard-model
gauge coupling constants.
That is, the larger dynamical scale 
$\Lambda \simeq 10^{13}-10^{14}~\GEV$ compared with the usual QCD scale
$\Lambda_{\rm QCD} \simeq 0.1~\GEV$ is a natural consequence of the
large number of colors $N_c = 6-10$.

\section*{Acknowledgments}

Y.N. thanks the Japan Society for the Promotion of Science for financial 
support.
This work is supported in part by the Grant-in-Aid, Priority Area
``Supersymmetry and Unified Theory of Elementary Particles''(\# 707).

\newpage

%
%
%
\newcommand{\Journal}[4]{{\sl #1} {\bf #2} {(#3)} {#4}}
\newcommand{\PL}{\sl Phys. Lett.}
\newcommand{\PR}{\sl Phys. Rev.}
\newcommand{\PRL}{\sl Phys. Rev. Lett.}
\newcommand{\NP}{\sl Nucl. Phys.}
\newcommand{\ZP}{\sl Z. Phys.}
\newcommand{\PTP}{\sl Prog. Theor. Phys.}
\newcommand{\NC}{\sl Nuovo Cimento}
\newcommand{\MPL}{\sl Mod. Phys. Lett.}
\newcommand{\PRep}{\sl Phys. Rep.}

\end{document}